\documentclass{moriond}

\def\babar{\mbox{\slshape B\kern-0.1em{\smaller A}\kern-0.1em
    B\kern-0.1em{\smaller A\kern-0.2em R~}}}

\def\be{\begin{equation}}
\def\ee{\end{equation}}
\def\bea{\begin{eqnarray}}
\def\eea{\end{eqnarray}}

\newcommand{\Photo}{}

\begin{document}
\vspace*{4cm}
\title{Determination of $\vert V_{ub} \vert$ at Belle II}

\author{M. Lubej on behalf of the Belle II Collaboration}

\address{Experimental Particle Physics Department, Jo\v zef Stefan Institute, Jamova cesta 39,\\
1000 Ljubljana, Slovenia}

\maketitle\abstracts{
Semileptonic $B$ meson decays involving low-mass charged leptons $e$ or $\mu$ are expected to be free of non-Standard Model contributions and therefore play a critical role in determinations of $\vert V_{ub} \vert$ and $\vert V_{cb} \vert$. Of all the CKM matrix parameters, $\vert V_{ub} \vert$ is the least precisely measured and in most need of additional studies in order to better constrain the apex of the Unitarity Triangle. We focus on exclusive reconstruction of charmless semileptonic $B$ meson decay $B \to \pi \ell \nu$, and present prospects and estimates for measuring $\vert V_{ub} \vert$ at Belle II with the full planned dataset of $50~\mathrm{ab}^{-1}$ of integrated luminosity.}

\section{Introduction}

Precise measurements of CKM matrix elements are necessary to probe the quark mixing mechanism of the Standard Model (SM) \cite{PhysRevLett.10.531,Kobayashi:1973fv} and to search for possible physics beyond it. Semileptonic decays of $B$ mesons involving low-mass charged leptons $e$ or $\mu$ are expected to be free of non-SM contributions and therefore play a crucial role in precise determinations of CKM matrix elements $\vert V_{ub} \vert$ and $\vert V_{cb} \vert$. Eq. \ref{eq:pseudo} represents the differential branching fractions for $B$ meson decays to pseudoscalar\footnote{This form has been simplified for low mass charged leptons $e$ and $\mu$.} states
\begin{equation}
\frac{\mathrm{d}\mathcal{B}(B \rightarrow P \ell \nu)}{\mathrm{d}q^2} = \vert V_{ub} \vert^2\frac{G_F^2 \tau_B }{24\pi^3} p_{P}^3 \vert f_{+}^{P}(\mathrm{q}^2) \vert{}^2,
\label{eq:pseudo}
\end{equation}

where $q^2$ is the squared value of the momentum transferred to the lepton pair, $f_{+}^{P}$ is one of the pseudoscalar form-factors, $\tau_B$ and $p$ the lifetime and momentum magnitude of the $B$ meson and the pseudoscalar, respectively, and $G_F$ the Fermi constant. From Eq. \ref{eq:pseudo} it can be seen that in order to determine the value of the $\vert V_{ub} \vert$ element, one needs a good measurement of the branching fraction and a good understanding of the form-factors that come into play. 

Of all the CKM matrix elements, $\vert V_{ub} \vert$ is the least precisely known due to the limited theoretical understanding of the form-factors and challenges with experimental measurements, so additional studies are needed in order to constrain the apex of the Unitarity Triangle even further. With Belle II we are entering an era of precision measurements, where we will be able to determine the value of $\vert V_{ub} \vert$ with precision at the percent level \cite{b2tip:2017}.

The most commonly studied decay mode to determine the magnitude of the CKM matrix element $\vert V_{ub} \vert$ is $B \to \pi \ell \nu$, because it offers a clean experimental measurement of the branching fraction and a precise theoretical calculation of the $B \to \pi$ form-factors. One of the problems with existing determinations of $\vert V_{ub} \vert$, the so-called $V_{ub}$ puzzle, is a persistent discrepancy between $\vert V_{ub} \vert$ measurements based on inclusive and exclusive charmless $B$ meson decays. In inclusive measurements we focus on all decays of the form $B \to X_u \ell \nu$ at the same time and try to reconstruct them inclusively, and in exclusive measurements we focus on specific decays, as in this case. 

\section{Reconstruction methods at $B$ factories}
\label{sec:rec}

Branching fraction measurements of leptonic and semileptonic $B$ mesons are possible using several different experimental techniques that differ only in the way how the companion $B$ meson is reconstructed. In this review we will only focus on the untagged (inclusive tag) method or the tagged method with hadronic tag decays.

In untagged analyses \cite{Ha:2010rf} we first reconstruct our signal $B$ meson, with the exception of the escaped neutrino. Since the detector hermetically covers a relatively large portion of the full solid angle, we can inclusively determine the 4-momentum of the companion $B$ meson by adding up the 4-momenta of all the remaining charged tracks and neutral clusters in the event, as shown in Eq. \ref{eq:roecalc}.
\begin{equation}
\label{eq:roecalc}
\mathrm{p}_{B_{comp}} = \sum_{i~=~{\rm tracks~and~clusters}}^{\rm rest~of~event} \left( E_i,\mathbf{p}_i\right).
\end{equation}
Due to the well known initial $\Upsilon(4S)$ state, we can determine the missing 4-momentum as
\begin{equation}
\label{eq:misscalc}
\mathrm{p}_{miss} = \mathrm{p}_{\Upsilon(4S)} - \mathrm{p}_{B_{sig}} - \mathrm{p}_{B_{comp}},
\end{equation}
which is equal to the 4-momentum of the missing neutrino, if neutrino is the only missing particle in the event. The untagged method has an efficiency of about $\mathcal{O}(10~\%)$, so it is the recommended approach with smaller data samples. The downside is the lower $q^2$ resolution, since 4-momentum of the companion $B$ meson is determined in an inclusive way.

In tagged analyses with the hadronic tag reconstruction \cite{Sibidanov:2013rkk} we first fully reconstruct the companion $B$ meson in several hadronic decay modes. After having a good companion $B$ meson candidate, we require that the rest of event is consistent with the signature of the signal decay. The missing 4-momentum can be imposed in the same way as in Eq. \ref{eq:misscalc}, taking into account the different method for $\mathrm{p}_{B_{comp}}$ calculation. Knowing exactly how the companion $B$ meson was reconstructed, this enables us to calculate $q^2$ with better precision, however, the efficiency drops to about $\mathcal{O}(0.1~\%)$, so this method is better suited for abundant data samples. 

\section{Belle II prospects}

Luminosity at the Belle II experiment is to be increased by a factor of $40$ compared to that of Belle, and we hope to acquire $50~{\rm ab}^{-1}$ of data, $50$ times more than in the case of its predecessor. In addition to more data, the detector itself is being upgraded, with improvements in almost all aspects, such as tracking and particle identification. Improvements were also made on the software side, with smarter and more precise algorithms. In order to study the impact of Belle II in the future, we are able to make some predictions using Monte Carlo (MC) simulations. The MC data sample used in these studies was a mixture of generic $B \overline B$ events and continuum $q \overline q$ events, where $q=u,~d,~s,~c$, which were produced in a Belle II Monte Carlo campaign in 2015 and correspond to a size of $500~{\rm fb}^{-1}$.

In case of the tagged MC study, the most relevant improvement, besides the detector upgrade, is the better tagging algorithm with significantly higher tagging reconstruction efficiency. From the MC study it was concluded that the overall reconstruction efficiency for $B \to \pi \ell \nu$ is about $0.55~\%$ \cite{b2tip:2017}, which is significantly higher than the overall reconstruction efficiency of $0.3~\%$ of the tagged measurement reported by Belle \cite{Sibidanov:2013rkk}. 

\begin{figure}
\begin{minipage}[t]{.48\textwidth}
\centering
  \includegraphics[width=0.8\textwidth]{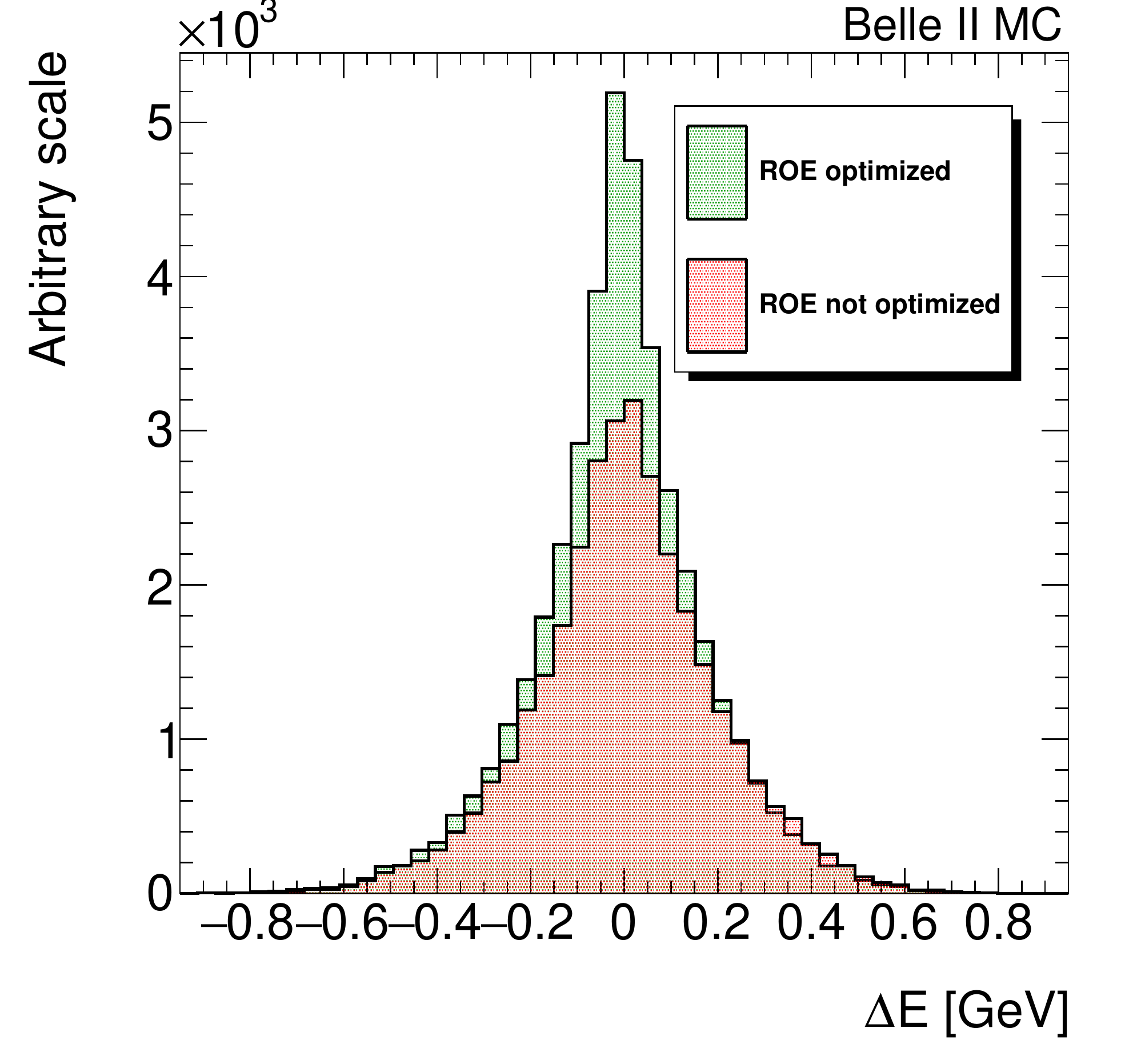}
\caption{The $\Delta E$ distribution of signal $B \to \pi \ell \nu$ candidates before and after rest of event clean-up. $\Delta E$ is a variable which compares the $B$ meson energy to the CMS energy of the initial beam. Signal candidates peak at zero.}
\label{fig:de}
\end{minipage}
\hfill
\begin{minipage}[t]{.48\textwidth}
\centering
  \includegraphics[width=0.8\textwidth]{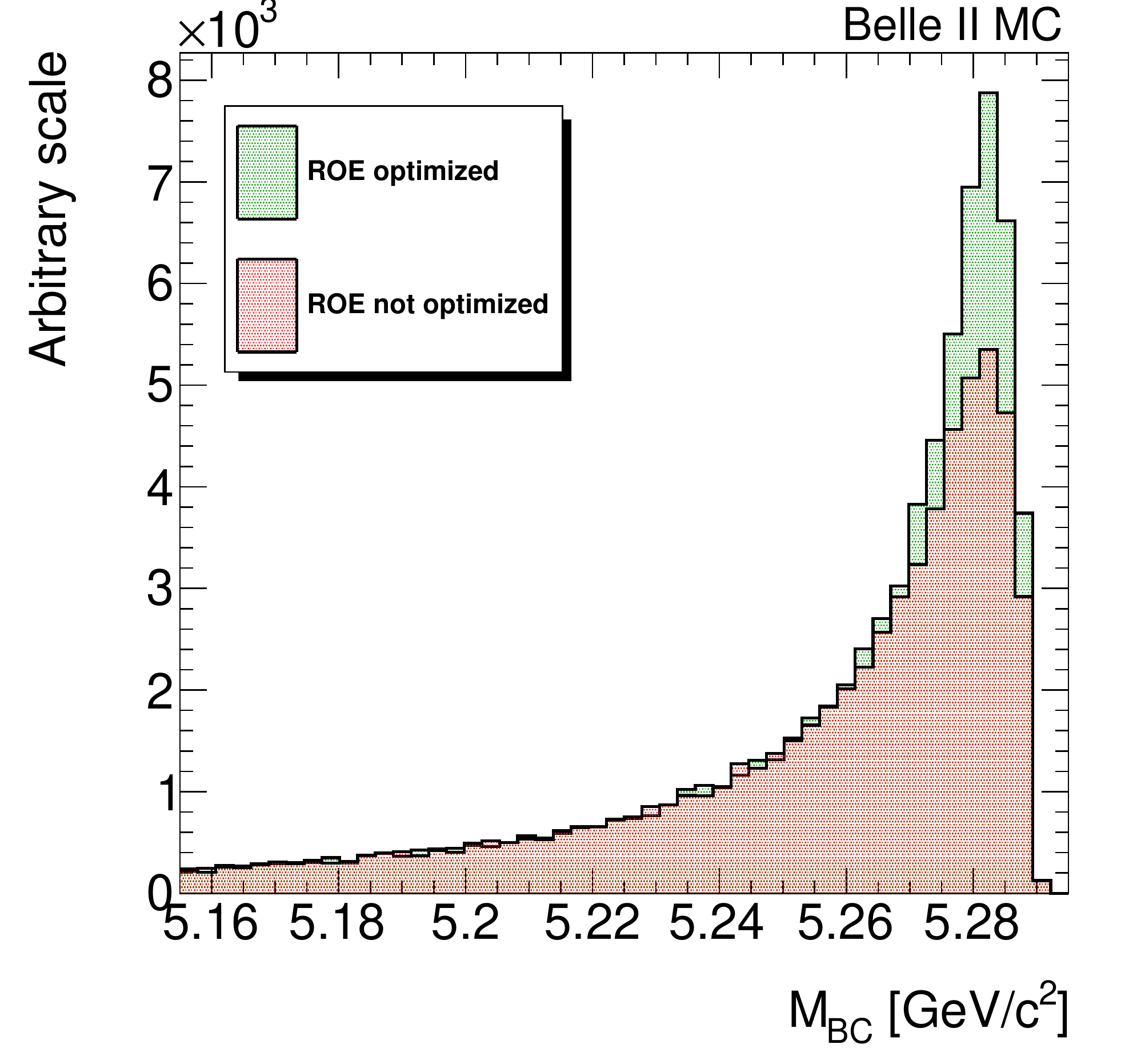}
\caption{The $M_{BC}$ distribution of signal $B \to \pi \ell \nu$ candidates before and after rest of event clean-up. $M_{BC}$ is a variable which constrains the $B$ meson to the CMS energy of the initial beam. Signal candidates peak at nominal $B$ meson mass with a narrow width.}
\label{fig:mbc}
\end{minipage}
\end{figure}

In case of the untagged MC study, improvements were made in the treatment of the remaining tracks and clusters in the event. Studies showed that the rest of event consists of a considerable amount of tracks and clusters which should not be taken into account. These objects, dubbed as extra tracks and clusters, are for example produced in secondary interactions of primary particles with detector material, or are fake candidates stemming from imperfect reconstruction. Either way, such cases should be discarded from our selection by cleaning up the rest of event. Taking all improvements into account, the overall reconstruction efficiency for this channel is around $20~\%$ \cite{b2tip:2017}, as opposed to $11~\%$ reported in the untagged measurement of $B \to \pi \ell \nu$ decays performed by Belle \cite{Ha:2010rf}. Figures \ref{fig:de} and \ref{fig:mbc} show the improvement effect of the rest of event clean-up on Belle II MC on the $\Delta E$ and $M_{BC}$ distributions, respectively.

Even though these studies were performed on a smaller MC sample, it is possible to extrapolate the errors to larger values of integrated luminosity, where one should keep in mind that the total error has contributions which do not scale with increasing data. Eq. \ref{eq:scaling} shows the general formula for error scaling with separate contributions to the total error,

\begin{equation}
\sigma_{\rm tot}(\mathcal{L}) = \sqrt{(\sigma_{\rm stat}^2(\mathcal{L}_0)+\sigma_{\rm sysred}^2(\mathcal{L}_0))\times\frac{\mathcal{L}_0}{\mathcal{L}}+\sigma_{\rm sysirred}^2(\mathcal{L}_0)},
\label{eq:scaling}
\end{equation}

where $\mathcal{L}$ is the arbitrary integrated luminosity which we are interested in, $\mathcal{L}_0$ is the integrated luminosity at which the initial studies/measurements were made, $\sigma_{\rm stat}$ is the statistical error, $\sigma_{\rm sysred}$ the reducible part and $\sigma_{\rm sysirred}$ the irreducible part of the systematic error. Based on the tagged and untagged Belle analyses at $711~{\rm fb}^{-1}$ \cite{Sibidanov:2013rkk} and $605~{\rm fb}^{-1}$ \cite{Ha:2010rf}, it is possible to estimate the precision limit of the irreducible systematic error to $2.0~\%$ and $1.6~\%$ \cite{b2tip:2017}, respectively.

\begin{figure}
\begin{minipage}[t]{.48\textwidth}
\centering
\includegraphics[height=0.8\textwidth]{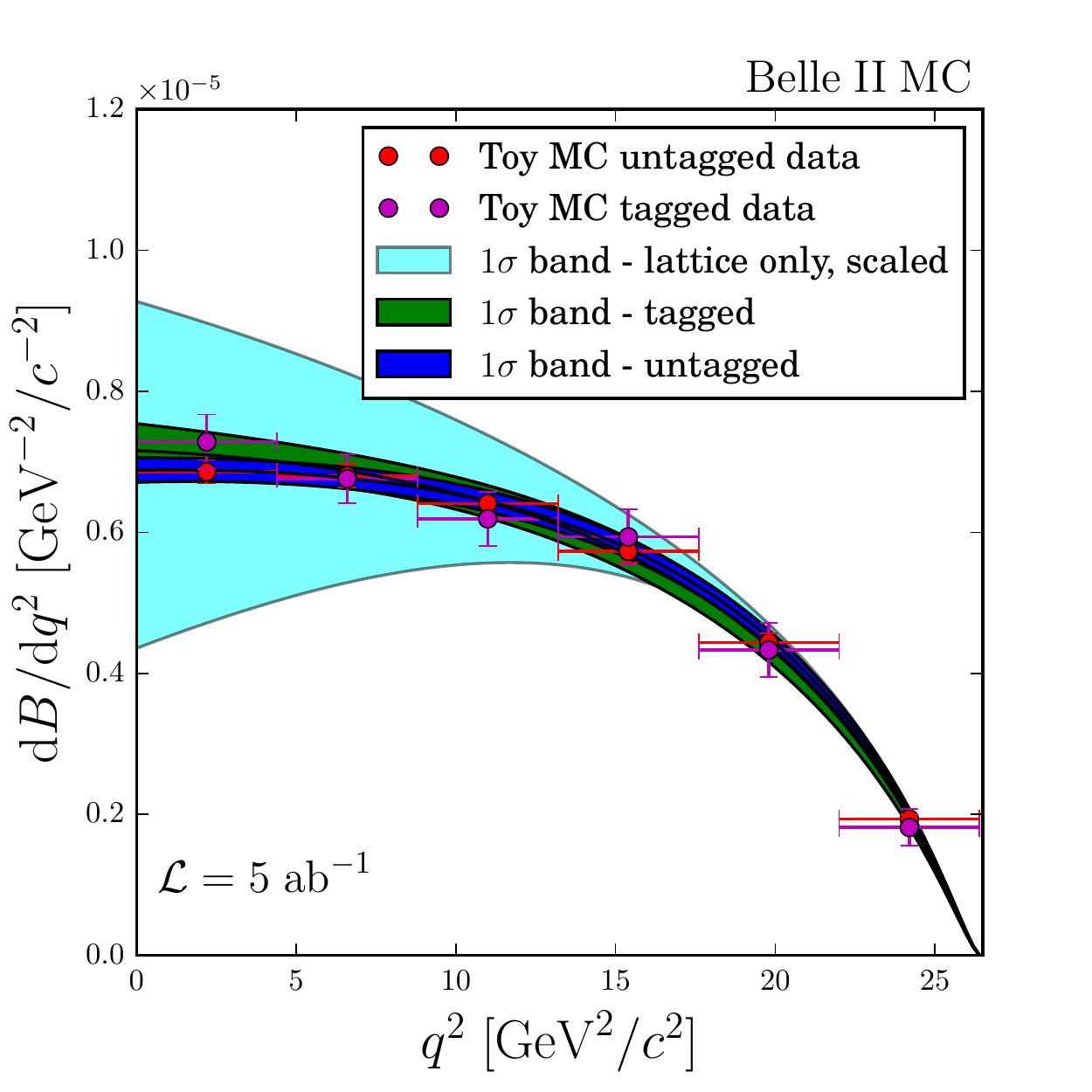}
\caption{An example of a Belle II simultaneous fit to MC data and LQCD input at $5~{\rm ab}^{-1}$ including LQCD forecasts in 5 years for the tagged and untagged analysis.}
\label{fig:vubFit}
\end{minipage}
\hfill
\begin{minipage}[t]{.48\textwidth}
\centering
\includegraphics[height=0.8\textwidth]{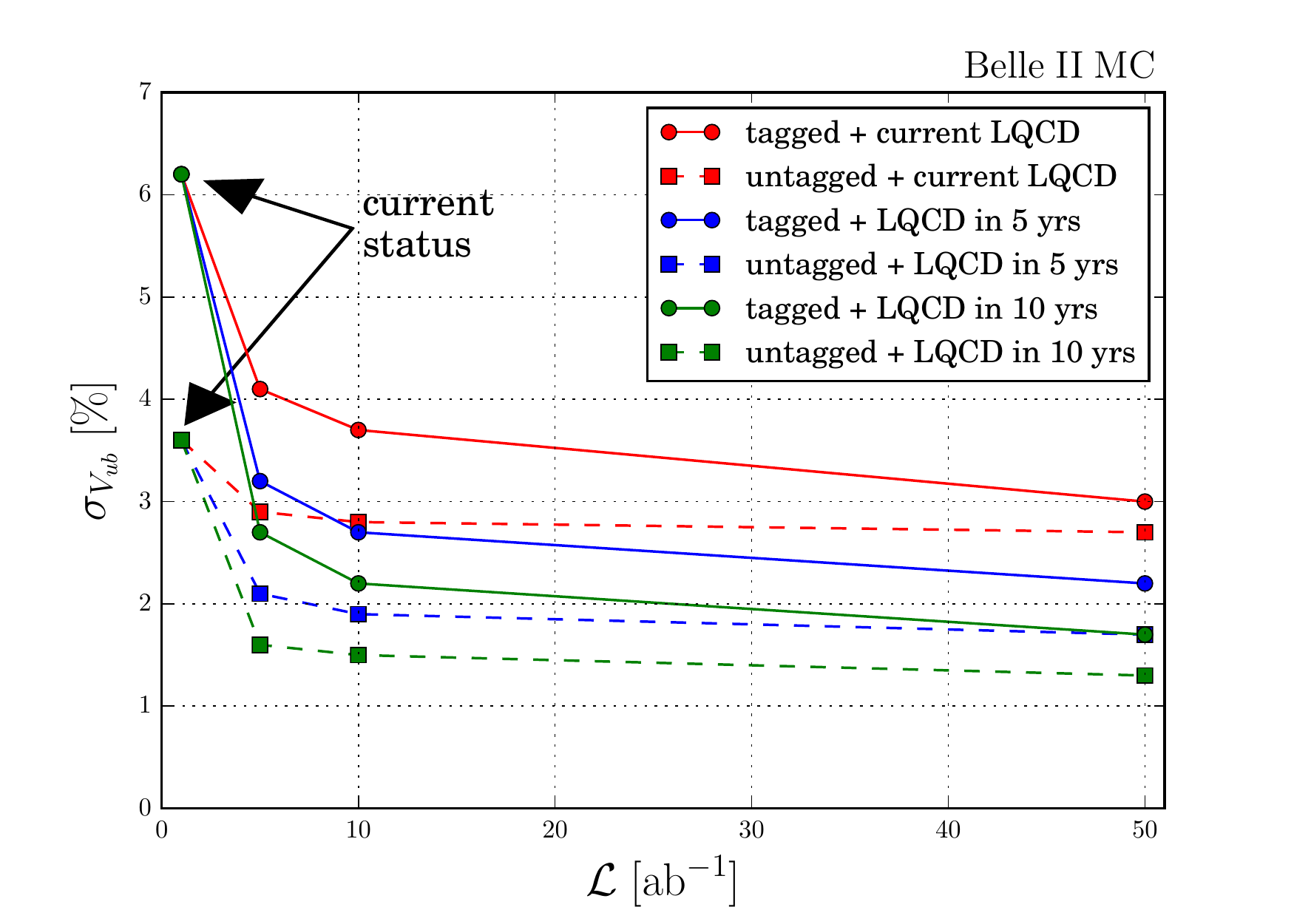}
\caption{$\vert V_{ub} \vert$ precision estimates for the tagged and untagged reconstruction method at $5,~10$ and $50~{\rm ab}^{-1}$ of integrated luminosity for current LQCD and LQCD forecasts in 5 and 10 years.}
\label{fig:vubErr}
\end{minipage}
\end{figure}

The $\vert V_{ub} \vert$ parameter was extracted from a simultaneous fit to the branching fraction distributions and LQCD input. The LQCD input used were the average form factor parameters of \cite{Lattice:2015tia} and \cite{Flynn:2015mha}, performed by \cite{Aoki:2016frl}. Optimally, LQCD input will become more precise as the Belle II detector acquires data, so it is possible to take LQCD error forecasts into account \cite{b2tip:2017,kronfeld_etal:2016}. Figure \ref{fig:vubFit} shows an example of a fit at $\mathcal{L} = 5~{\rm ab}^{-1}$, while Figure \ref{fig:vubErr} shows the precision of $\vert V_{ub} \vert$ for multiple values of integrated luminosity from $5$ to $50~{\rm ab}^{-1}$ for current LQCD and LQCD forecast in the next decade \cite{b2tip:2017}. With the full Belle II dataset and the LQCD forecasts in 10 years we estimate the precision on $\vert V_{ub} \vert$  to be $1.7~\%$ and $1.3~\%$ in the case of the tagged and untagged analysis, respectively.

\section{Conclusions}

In this paper we present prospects on $\vert V_{ub} \vert$ CKM matrix parameter measurements with the full planned Belle II data sample using detailed MC studies. We estimate the precision to at least a $1.7~\%$ and $1.3~\%$ in the case of a tagged and untagged reconstruction method of the companion $B$ meson, respectively.

\section*{Acknowledgments}

I would like to congratulate the organizers for another excellent conference with very interesting contributions and great hospitality.

Additionally, I would also like to thank the Belle II Speakers Committee for giving me the opportunity to represent the Belle II collaboration at such a prestige conference.


\section*{References}

\begin{thebibliography}{1}

\bibitem{PhysRevLett.10.531}
Nicola Cabibbo.
\newblock Unitary symmetry and leptonic decays.
\newblock {\em Phys. Rev. Lett.}, 10:531--533, Jun 1963.

\bibitem{Kobayashi:1973fv}
Makoto Kobayashi and Toshihide Maskawa.
\newblock {CP Violation in the Renormalizable Theory of Weak Interaction}.
\newblock {\em Prog. Theor. Phys.}, 49:652--657, 1973.

\bibitem{b2tip:2017}
{The Belle II collaboration and B2TiP theory community}.
\newblock {\em The Physics Prospects for Belle II}.
\newblock To be published in 2017.

\bibitem{Ha:2010rf}
H.~Ha et~al.
\newblock {Measurement of the decay $B^0\to\pi^-\ell^+\nu$ and determination of
  $|V_{ub}|$}.
\newblock {\em Phys. Rev.}, D83:071101, 2011.

\bibitem{Sibidanov:2013rkk}
A.~Sibidanov et~al.
\newblock {Study of Exclusive $B \to X_u \ell \nu$ Decays and Extraction of
  $\vert V_{ub}\vert$ using Full Reconstruction Tagging at the Belle
  Experiment}.
\newblock {\em Phys. Rev.}, D88(3):032005, 2013.

\bibitem{Lattice:2015tia}
Jon~A. Bailey et~al.
\newblock {$|V_{ub}|$ from $B\to\pi\ell\nu$ decays and (2+1)-flavor lattice
  QCD}.
\newblock {\em Phys. Rev.}, D92(1):014024, 2015.

\bibitem{Flynn:2015mha}
J.~M. Flynn, T.~Izubuchi, T.~Kawanai, C.~Lehner, A.~Soni, R.~S. Van~de Water,
  and O.~Witzel.
\newblock {$B \to \pi \ell \nu$ and $B_s \to K \ell \nu$ form factors and
  $|V_{ub}|$ from 2+1-flavor lattice QCD with domain-wall light quarks and
  relativistic heavy quarks}.
\newblock {\em Phys. Rev.}, D91(7):074510, 2015.

\bibitem{Aoki:2016frl}
S.~Aoki et~al.
\newblock {Review of lattice results concerning low-energy particle physics}.
\newblock 2016.

\bibitem{kronfeld_etal:2016}
Private communications with \href{mailto:ask@fnal.gov}{A. Kronfeld},
  \href{mailto:silvano.simula@roma3.infn.it}{S. Simula} and
  \href{mailto:takashi.kaneko@kek.jp}{T. Kaneko}, Oct. 15, 2016.

\end{thebibliography}
\end{document}